\documentclass[10pt,letterpaper]{article} 
\usepackage[english]{babel}
\newcommand{\cals}{{\cal S}}
\newcommand{\calsprime }{{{\cal S}^\prime}}

\newcommand{\calr}{{\cal R}}

\newcommand{\calk}{{\cal K}} 
\newcommand{\calh}{{\cal H}} 
\newcommand{\cala}{{\cal A}} 
 
\newcommand{\lb}{\langle}
\newcommand{\rb}{\rangle}

\newcommand{\beq}{\begin{equation}}
\newcommand{\eeq}{\end{equation}}
\newcommand{\lbl}{\label}
\newcommand{\beqnar}{\begin{eqnarray}}
\newcommand{\eeqnar}{\end{eqnarray}}
\newcommand{\beqnars}{\begin{eqnarray*}}
\newcommand{\eeqnars}{\end{eqnarray*}}

\newcommand{\s}{\\[1ex]}
\newcommand{\re}[1]{(\ref{#1})}
\newcommand{\q}{\quad}

\newcommand{\tnr}{\otimes} 
\newcommand{\tr}{\mbox{tr }}

\newcommand{\Sprime}{S^\prime }

\newtheorem{proposition}{Proposition}
\newtheorem{theorem}[proposition]{Theorem}

\newcommand{\boxx}{\rule{1mm}{2mm}}

\newcommand{\ts}{{\cals \tnr \cala}}

\begin{document} 
%
\begin{center}
\Large
\bf
Quantum measurements need not conserve energy: relation to
the Wigner-Araki-Yanase theorem
\\
\large
\rm 
\ 
\\
Stephen Parrott%
\footnote{For contact information before mid-October, 2016, 
go to http://www.math.umb.edu/$\sim$sp.  I may be unavailable for
an indeterminate period after that, so please do not be offended 
if emails are unanswered.
}
\ 
\s
\normalsize 
October 10, 2016
\end{center}
\begin{abstract}
The paper focuses on the fact that quantum projective measurements
do not necessarily conserve energy. On the other hand the 
Wigner-Araki-Yanase (WAY) theorem states that assuming a ``standard''
von Neumann measurement model and ``additivity'' of the total 
energy operator, 
projective measurements of a system {\em must}
conserve energy as defined by the system's energy operator. 
This paper explores the ideas behind  the WAY theorem in hopes of 
uncovering the origin of the contradiction. 

After Araki and Yanase published their proof of the WAY theorem,
Yanase appended a new condition now known as the Yanase condition.  
Under the simplifying assumption that the observable  
being measured has discrete and non-degenerate eigenvalues,
we prove that the Yanase condition actually follows from the hypotheses
of the original WAY theorem.  

The paper also proves that the hypotheses of the WAY theorem, together with
 the simplifying assumption, imply that 
the energy operator for the measuring apparatus
must be a multiple of the identity, which seems physically unlikely.
It seems probable that this surprising conclusion, along with the
Yanase condition,  also holds without the simplifying assumption.      
\end{abstract}
\section{Introduction}
In orthodox quantum mechanics, there are two ways that a quantum system
can change:  
\begin{enumerate}
\item
Continuous evolution as described by the Schroedinger equation;
\item
A discrete change (``collapse of the wave function'') caused by
a measurement of the system.

\end{enumerate} 
For continuous evolution, conservation of energy is automatically
enforced by the mathematical structure of the theory.  If the (mixed)
state of the system at time $t = 0$ is $\rho(0)$ and the Hamiltonian (energy
operator) is $H$, then the state $\rho(t)$ at time $t$ is 
$$
\rho(t) = e^{-iHt} \rho(0) e^{iHt}  .  
$$
At a given time, 
the (average) energy of state $\sigma$ is $\tr[ H \sigma] = \tr[\sigma H] $, 
where
$\tr$ denotes trace, so the energy of $\rho(t)$
is 
\begin{eqnarray*} 
\lefteqn{\mbox{energy of $\rho(t)$} = \tr[e^{-iHt} \rho(0)e^{iHt} H]} \\ 
&=& \tr[\rho(0) e^{iHt}H e^{-iHt}] = \tr[ \rho(0)H] = \mbox{energy of $\rho(0)$}. 
\end{eqnarray*}

Since conservation of energy is one of the most universal and cherished 
principles of physics, one might expect a similar trivial calculation to
establish conservation of energy for the discrete transition caused by 
a measurement.  But such is not the case. 

For our purposes, a measurement is described by a finite collection of 
``measurement operators'' $M_1, M_2, \ldots, M_n$. 
If the state of the system before measurement is $\rho$, the 
state after the measurement (disregarding the measurement result,
or assuming it unknown) is%
\footnote{This may seem strange because normally, one might suppose 
that after a measurement, the result is known!  
See Subsection \ref{review} for an explanation.
}
\footnote{This is the formulation of the standard text \cite{n/c}, which
is not the most general formulation.  For a clear account of the 
general formulation, see Chapter 1 of Jacobs' book \cite{jacobs}.  
For projective measurements, which is all that we need consider,
the two formulations are equivalent.
}

\beq
\lbl{eq10}
\sum_i M_i \rho M^\dag_i \q.
\eeq

For the measurement to conserve (average) energy for all
states $\rho$, it is necessary and sufficient that for all $\rho$,
\beq
\lbl{eq15}
\sum_i \tr[M_i\rho M^\dag_i H] = \tr[H\rho] \q,
\eeq
which, using the cyclic property of the trace, is equivalent to 
\beq
\lbl{eq20}
\sum_i \tr[(M^\dag_i H M_i -H)\rho)]\q.
\eeq
It is plausible, and a simple exercise%
\footnote{For $\rho = P_\phi$, the projector on a pure state $\phi$,
and any operator $K$, $\tr[KP_\phi] = \lb \phi, K\phi \rb$, and
it is a standard fact that $\lb \phi, K\phi \rb = 0 $ can hold for all
$\phi$ only for $K = 0$. 
}
to prove, that
this will hold for all states $\rho$ if and only if,
\beq
\lbl{eq30}
\sum_i M^\dag_i H M_i = H \q.
\eeq
In summary, only the special measurements satisfying \re{eq30}
can conserve energy.  

For projective measurements (i.e., the $M_i$ are orthogonal projectors
which sum to the identity), readers familiar with operator theory may
recognize that \re{eq30} is equivalent to requiring that the $M_i$
commute with $H$.  This simpler condition underscores the very special
nature of energy-conserving measurements.  
\\[2ex]
{\bf History}

It has been recognized at least since a seminal 1952 paper of Wigner
\cite{wigner} that some quantum measurements do not conserve energy.
Wigner's observations were generalized ten years later by  
Araki and Yanase \cite{a/y} in what has become known as 
the WAY (Wigner-Araki-Yanase) theorem, on which there is an extensive
modern literature.%
\footnote{ 
Of which I have had time to read only a small fraction since I learned 
of the WAY theorem a few months ago.  
(See, for example, the bibliography of \cite{l/b}.)
If I have overlooked some reference which should be included,
that is the reason.
}

I first learned of the possibility that quantum measurements might not
conserve energy from the charmingly written \cite{e-r}, which presents
a particular example of this phenomenon.  It does not mention the WAY
theorem, of which I assume its author was unaware, as was I.  
At that time, I formulated the core of the present work, but did not write
it up because it is so mathematically trivial that I assumed that it
must be known.   On learning from references
like \cite{popescu, ahmadi, l/b}   
that the WAY theorem is of current
interest, I thought that perhaps those observations might
be of some interest, despite their mathematical simplicity.

The informal remarks of the Introduction essentially constitute a 
proof of a theorem stating 
that a quantum projective measurement 
conserves energy if and only if its projectors commute with the
energy operator.  This is formally stated as Theorem \ref{thm1} 
in Section \ref{energythm}.  

 When I learned of the WAY theorem in the last few months, 
of course I wondered what might be its relation to 
the simple Theorem \ref{thm1}.   
The conclusion of the WAY theorem 
implies that {\em all} discrete quantum mechanical
observables%
\footnote{More precisely, all observables which can be ``exactly''
measured, which according to the usual textbook quantum mechanics,
means all observables.  The seminal papers on the WAY theorem 
also discuss a more general notion of ``approximate measureability''.
}
must 
commute with the energy observable! If physically correct, 
this would probably
destroy much of the structure of quantum mechanics and its explanatory 
power.  That makes it hard to believe. 

But notice that I said the {\em  conclusion} of the WAY theorem.
Of course, the WAY theorem has hypotheses, which include acceptance of
a so-called ``standard'' measurement model of von Neumann.  
Another important hypothesis assumes that the energy observable
is of a particular ``additive'' form.
If the WAY theorem's conclusion is unbelievable, 
then  chances are that one of its hypotheses is physically unrealistic. 

In a search for the origin of the conundrum
I studied a measurement model similar to the ``standard'' model,
but algebraically more natural. 
It also led to WAY-type theorems
with the same physical difficulties as the original WAY theorem.  
But the proofs seemed simpler and more
transparent.  
The new WAY-type theorems will be presented in Section \ref{waytype}.
The proof of the original WAY theorem (under the simplfying assumption
that the system observable has non-degenerate eigenvalues)
appears as a simple corollary in Section \ref{traditional},
which discusses the relations between the ``standard'' measurement
model used by Araki and Yanase \cite{a/y} and the more general 
model introduced in Section \ref{waytype}.  

The reader will naturally wonder if 
the time invested in working through this material will be adequately 
repaid by the understanding gained.  In honesty, I feel compelled 
to let him%
\footnote{
Or her, {\em of course}.  I adhere to the long-standing and sensible
grammatical convention that in contexts like this, ``him'', ``her'',
``him or her'', and ``her or him'' carry identical meanings.    
}
know that it may not be, unless he is already a connoisseur of the
ideas surrounding the WAY theorem.
I have been led to regard 
von Neumann type measurement theory in general and the WAY theorem's 
place in it as of questionable physical relevance.

On the other hand, the WAY theorem is characterized as ``famous'',
``remarkable'', and ``important'' in three recent papers by different authors, 
which makes me wonder if the difficulties which I have noticed are 
widely known.
In retrospect, I regret the time I have invested in studying it,
but having made the investment, it seemed worthwhile to write down
what I have learned to save others the trouble.   
\\[2ex] 
{\bf Notation:}
\\[2ex]
I hope that the notation informally introduced above will seem reasonably
natural 
to most readers.  We will not use  the Dirac notation 
$| s \rb$ for pure states, instead using the simpler $s$ 
which will be defined explicitly in the text instead of implicitly by 
Dirac notation.%

The advantage of Dirac notation is that when one sees $| s \rb$ 
(in browsing through a journal, say),
one immediately knows what it represents.  But Dirac notation is messy and 
sometimes hard to parse in more complicated expressions.
Instead of the Dirac
$| s \rb \lb s |$ to represent the projector on the pure state $s$, we shall
use the simpler $P_s$.   In contexts in which 
the pure state $s$ is considered as the mixed 
state (positive operator of trace 1) which is $P_s$, 
we sometimes write $\tilde{s}$ instead of $P_s$.    

We do not distinguish operators with carets, writing, 
for example,
$M_i$ instead of $\widehat{M_i}$.
The identity operator will be denoted $I$, with the space on 
which it acts determined by the context.  

\section{Energy-conserving projective measurents must commute 
with the energy operator}
\lbl{energythm}
We start with equation \re{eq30} in the Introduction, 
$$
\sum_i M^\dag_i H M_i = H \q, \eqno (\ref{eq30})
$$
which is equivalent to energy conservation for all states 
and show that when the $M_i$ are orthogonal projectors which sum to 1
(that is, the measurement is a projective measurement),
they all must commute with $H$.  

This holds for any operator 
$H$ representing a quantity conserved by the measurement,  
not just the energy operator.  The Hilbert space on which
$H$ operates can be finite or infinite dimensional.  The collection 
$\{M_i\}$ can be finite or infinite.

To emphasize that the $M_i$ are assumed to be projectors 
(usually called ``projection operators'' or simply ``projections'' 
in the mathematical literature), 
we write $P_i$ instead of $M_i$.  Thus the $P_i$
satisfy
$$
P^\dag_i = P_i \q, \q P^2_i = P_i \q \mbox{for all $i$,} \q 
P_i P_j = 0 = P_j P_i \q \mbox{for $i \neq j$, and }  \ \sum_i P_i = I .
$$ 

\begin{theorem}
\lbl{thm1}
Let $H$ be a given operator, and $\{P_i\}$ a collection of orthogonal
projectors which sum to the identity.  If the measurement defined 
by $\{ P_i \}$ conserves the observable quantity corresponding to $H$ 
(i.e., \mbox{if $\sum_i P_i H P_i = H$}) then 
all the $P_i$ commute with $H$:  $P_i H = H P_i$ for all $i$.
\end{theorem}  
{\bf Proof:}
\\[1ex]
The Introduction explained what it means for a measurement to ``conserve
energy'', and the same definition is used for any observable $H$ instead 
of energy.  Here it means that 
$$
\sum_i \tr [ P_i H P_i \rho ] = \tr [H \rho] \q 
\mbox{for all mixed states $\rho$,} 
$$
which is equivalent to 
$$
\sum_i P_i H P_i =  H \q.
$$
For any $k$, 
$$
P_k H = \sum_i P_k P_i H P_i = P_k H P_k 
$$
because $P_k P_i = 0$ for $i \neq k$ and $P^2_k = P_k$. Similarly,  
$$
H P_k = \sum_i  P_i H P_i P_k = P_k H P_k \q \mbox{so}
$$
$$
P_k H = P_k H P_k = H P_k .
\q \boxx
$$ 
Thus most projective measurements do not conserve energy; only
the very special ones that commute with the energy operator $H$ can.  

It is natural to wonder if the same or something similar is true
for general, non-projective measurements. Again, it it trivial that
if the measurement operators commute with $H$, then the measurement
conserves the physical quantity corresponding to $H$.  However, the 
converse is not so evident as for projective measurements, and may 
not be true.  

Mathematicians may be interested in this problem.  However     
its physical relevance is probably minor unless there turns out to
be a simple,
general condition that is equivalent to conservation of energy.  
The fact that typical projective measurements don't conserve energy 
already poses a problem for the foundations of quantum mechanics.

\section{Reviews}
\lbl{way} 
The reviews of this section are included to make the paper more 
nearly self-contained.  Many will have no need for much of it.
I suggest skimming and referring back to it when needed.  But please do 
read the first paragraph of Subsection \ref{review} for the definition
of {\em measurement} which will be used throughout. 

\subsection{Review of measurement operators}
\lbl{review}
A {\em measurement} in quantum mechanics is specified by a collection
$\{ M_i \}$ of {\em measurement operators} $M_i$.  For simplicity
of language, we shall often refer to the collection $\{ M_i \} $ 
of measurement operators as a {\em measurement}.%
\footnote{
The rest of the paper will deal exclusively with projective measurements,
so this section in unnecessarily general.  This occurred 
because the first section was written before it became clear that
the rest would not require the generality.  However, specializing
to projective measurements does not simplify anything, so I decided
not to reset the type.
}

A collection $\{M_i \}$ of   measurement operators is required to
satisfy 
$$
\sum_i M^\dag_i M_i = I .
$$
The index $i$ can run over any countable set, which when the set 
is finite is usually taken to 
be the set $\{1, 2, \ldots , N\}$ of the first $N$ integers, and this
is the only situation that we shall consider.  

The result of a measurement is one of the integers in this index set.
For a quantum system in mixed state $\rho$, the probability $p(i)$ 
that the measurement result is $i$ is 
$$
p(i) = \tr [ M_i \rho M^\dag_i ] \q,
$$
and the measurement changes the premeasurement state $\rho$ to
the postmeasurement state 
\beq
\lbl{eq40}
\frac{M_i \rho M^\dag_i}{ \tr [ M_i \rho M^\dag_i]} 
\q.
\eeq 
The denominator, necessary to normalize the trace to 1, is just $p(i)$. 
If we know that the measurement has been made but do not know the result,
then the postmeasurement state is the mixed state which is the weighted 
average of \re{eq40} with weights the probabilities $p(i)$:
\beq
\lbl{eq50} 
\mbox{postmeasurement state} = \sum_i p(i) 
\frac{M_i \rho M^\dag_i}{p(i)} = \sum_iM_i \rho M^\dag_i \q. 
\eeq

If the mixed state $\rho$ happens to be a pure state 
$\rho = \tilde{\phi} = P_\phi$,
then \re{eq40} is always pure (it is the normalization of $M_i \phi$),
but \re{eq50} is rarely pure.  Because measurement usually converts pure
states into non-pure mixed states, the language of mixed states 
(positive operators of trace 1) is more natural than the language of pure
states (unit vectors in a Hilbert space) to describe measurement operations.

Many classical papers are formulated in terms of 
``measurement of observables'', which is a slightly different kind of 
measurement.  An {\em observable} (like position, momentum, or energy)
is mathematically represented by a Hermitian operator on a Hilbert space.
For simplicity, we shall consider only observables whose spectrum consists
only of eigenvalues.  (The seminal papers which we shall discuss 
such as \cite{wigner, a/y, yanase}
also make this assumption.)  Let $\{\lambda_i\}$  be the collection of
distinct eigenvalues of an observable $H$, and let $P_i$ denote the projector
on the eigenspace for eigenvalue $\lambda_i$.  Then the $P_i$ are orthogonal
projectors (i.e., $P_i P_j = 0$ for $i \neq j$) which sum to the 
identity operator $I$, and the 
collection $\{P_i \}$ constitutes a special kind of measurement operators.  
A measurement made with orthogonal {\em projectors} which sum to the identity  
is called a {\em projective} measurement. 

Note that by definition, projectors are Hermitian operators. 
Also, the spectral theorem states that  
$$
\sum_i \lambda_i P_i = H \q.
$$ 

Suppose we have a large collection of identical states $\rho$ and perform 
a measurement on each one.  Each measurement yields a result $i$,
with which is associated an eigenvalue $\lambda_i$, physically 
interpreted as the measured value of the observable.  The average 
of all these measured values, for a  large enough sample, should be 
close to 
\beq
\lbl{eq60}
\sum_i p(i) \lambda_i = \sum_i \tr[P_i \rho P^\dag_i] \lambda_i 
= \sum_i \tr[\rho P^2_i] \lambda_i 
= \tr[\rho \sum_i P_i \lambda_i]  = \tr[\rho H].  
\eeq
Thus $\tr[\rho H]$ is the mathematical representation of the average value
obtained by measuring the observable many times on 
identical systems in state $\rho$.

Let $H$ be an observable which we shall call the energy observable for
ease of language, though what we say will apply to any observable such
as spin, etc. 
To say that a measurement $\{M_i\}$ {\em conserves} energy $H$
(or if $H$ is an observable other than energy, 
whatever quantity it represents)
means that for all mixed states $\rho$, the average energy 
of the postmeasurement state is the same as the average energy of 
the premeasurement state $\rho$, i.e., that
\beq
\lbl{eq70}
\tr[  \rho H ] = \sum_i \tr[ M_i \rho M^\dag_i H] \q.
\eeq  
As previously noted, this is equivalent to 
$$
\sum_i M^\dag_i H M_i = H \q. 
$$
\subsection{Review of isometries}
Let $\calh$ and $\calk$ be Hilbert spaces.  A linear transformation
$U: \calh \rightarrow \calk$ is called an {\em isometry} if it preserves
inner products (and consequently norms), that is, if
$$
\lb U \phi, U \psi \rb = \lb \phi, \psi \rb
\q \mbox{for all $\phi, \psi \in \calh$.}
$$
Such isometries $U$ are often incorrectly called ``unitary'' operators in 
the physics literature.  The difference between a unitary operator 
and an isometry is that a unitary operator is required to be surjective
(``onto''); i,e., the range of a unitary operator must be all of $\calk$.
This difference will be important to us, so it is worthwhile to 
note it explicitly.

It is routine to verify that $U$ is an isometry if and only if  
\beq
\lbl{eq73}
U^\dag U = I \q.
\eeq
One also easily checks that $UU^\dag$ is the projector on the range of 
$U$.  Thus a unitary operator satisfies in addition to \re{eq73},
$$
UU^\dag = I.
$$ 

For later reference, it is worth noting 
that, from \re{eq73}, the adjoint $U^\dag$ 
of an isometry $U$ acts as an inverse for $U$ on the range of $U$:  
$$
\mbox{If} \q U \phi = \psi,\q \mbox{then $U^\dag \psi = \phi$.} 
$$
Also, we shall use the following simple fact.  Let $R$ denote the 
projector on the range of $U$.  Then obviously,
$$
RU = U,  \q \mbox{and taking adjoints shows that also} \q U^\dag R 
= U^\dag . 
$$    
\subsection{Review of a von Neumann type measurement model} 
\lbl{model}

This subsection sets up the measurement model in which the WAY
theorem is formulated.     The model is generally  attributed 
to von Neumann.  
Busch and Lahti \cite{b/l} call it the ``Standard Model of Quantum
Measurement Theory''.  
Readers already familiar with the von Neumann model may only need to 
skim this section. 

In the early days of quantum theory, much attention was given to 
the transition between the classical world governed by everyday
Newtonian physics and the much stranger quantum world seemingly 
governed entirely differently.  Exactly how does the quantum world
become classical?

It seemed that light might be shed on this problem by examining in detail 
the process of measuring an observable like the spin of a spin-1/2
quantum particle in a given direction.
According to quantum theory, the measurement is a projective measurement
implemented by two projectors $\{P_+ , P_- \}$  corresponding to ``up''
and ``down'' spins. Some feel that this measurement occurs on a  
quantum level which has to be somehow amplified to be classically
observable.  For example, spin can be observed with a macroscopic 
Stern-Gerlach apparatus which seems to obey the laws of Newtonian physics.

I am trying to explain a point of view 
with which I have never been comfortable. 
I don't see why the Stern-Gerlach apparatus could not be regarded as 
a physical implementation of the measurement operators $\{ P_+, P_- \}$.

However, suppose we accept the interpretation that the $\{ P_+, P_- \}$
measurement has to somehow be amplified to be observable on the classical
level.  The following mechanism, usually attributed to von Neumann, 
has been proposed.

Let $s_+$ and $s_-$ be the (pure) quantum states which are eigenvectors
of $P_+$ and $P_-$: 
$$P_+ s_+ = (1/2) s_+, \q P_- s_- =  (- 1/2) s_-  \q.
$$
We imagine
that a Stern-Gerlach apparatus also has a complete set 
of two orthogonal pure states $a_+ , a_-$.  Consider the association
$$
s_+ \mapsto s_+ \tnr a_+ , \q  
s_- \mapsto s_+ \tnr a_-  \q.
$$  
Here $s_\pm \tnr a_\pm$ are states in a Hilbert space $\ts$ which is the tensor
product of the original state space $\cals$ for the particle 
whose spin is being
measured and a Hilbert space $\cala$ for the apparatus.  Then we perform the 
measurement with projection operators $I \tnr P_{a_\pm}$, with 
$\{P_{a_+} , P_{a_-} \}$ a projective measurement in the apparatus space.  The
postmeasurement state is then either $s_+ \tnr a_+$ or $s_- \tnr a_-$
(which after tracing out the state space becomes $a_+$ or $a_-$).

This process is supposed to somehow explain how quantum measurements
get converted to classical ones which we can perform in the laboratory.
To me, it seems rather silly, no more explanatory than simply imagining
the quantum measurement with $\{P_\pm \}$ as implemented in some way 
which we choose not to (or cannot) describe in detail.

Anyway, the WAY theorem is formulated in terms of this 
measurement model, so we have to tentatively accept it to continue.
The observable $S$ which we want to measure will be arbitrary, not
necessarily spin.  It operates on a Hilbert space $\cal S$. 
We shall call $S$ the {\em system observable} and $\cals$ the 
{\em system space}.  

The measuring apparatus (assumed to obey the laws of quantum 
mechanics despite the above motivating remards) is an observable on a 
Hilbert space $\cal A$.  The system together with apparatus operates
on $\cals \tnr \cala$.  

\section{WAY-type theorems in a new framework}
\lbl{waytype}
\setcounter{subsection}{-1}
\subsection{Warning}
\lbl{warning}
Subsection \ref{interpretation} presents a measurement theory which
is similar to the ``standard'' von Neumann theory, but not quite 
equivalent.  It is a little more general, and, I think,
algebraically more natural.  
Section \ref{traditional} makes contact with the traditional 
von Neumann measurement theory in which Araki and Yanase's original
proof \cite{a/y} of the WAY theorem is formulated.

\subsection{Interpretations of Araki and Yanase's setup}
\lbl{interpretation}
Following Araki and Yanase \cite{a/y}, we assume that the system 
observable $S$ has discrete spectrum (possibly degenerate) with 
distinct eigenvalues $\{ \lambda_i \}$.  Let $Q_i$ denote the projector
on the eigenspace for eigenvalue $\lambda_i$.  Then the spectral theorem
states that  $Q_iQ_j = 0$ for $i \neq j$ and   
\beq
\lbl{eq74}
S = \sum_i \lambda_i Q_i \q.
\eeq 

We shall actually be concerned with the projective measurement with
measurement operators $\{ Q_i \}$ rather than measuring $S$ itself; 
that is, the eigenvalues of $S$ will not enter into our considerations 
beyond the definition \re{eq74}.
Also, little insight will be lost by assuming that each $Q_i$ has one-dimensional
range spanned by a unit vector $\phi_i$, and 
we shall first consider this case when it simplifies the exposition.   
Our generalizations to $Q_i$ of arbitrary dimension (finite or infinite) 
will be routine. 

Let $\phi_i$ be an orthonormal basis for the system space 
$\cals$ such that the range of each $Q_i$ is spanned 
by some subcollection of $\{ \phi_i \}$, where $i$ runs over some 
index set..  
Corresponding to this orthonormal basis 
let $\{ X_i \}$ be an orthonormal basis , for the apparatus space $\cala$, 
where $i$ runs over the same
index set.%
\footnote{It might seem perverse to use Greek $\phi_i$ to denote vectors
in $\cals$ and Roman $X_i$ for $\cala$, but this is the notation of 
Araki and Yanase \cite{a/y}.  We use it both for comparison and also
because it seems to make it easier to sort out at a glance which vector
is in which space.} 
Let $U$ be the unique isometry $U: \cals \rightarrow \cals \tnr \cala$
satisfying 
\beq
\lbl{eq74.10} 
U \phi_i = \phi_i \tnr X_i \q \mbox{for all $i$}
\q.
\eeq
Here we depart from the ``standard''  
von Neumann measurement model, and from Araki and Yanase \cite{a/y}
in particular.  The differences may seem small, but the setups are 
not equivalent as one might imagine.%
\footnote{When this subsection was written, I thought that they
{\em would} turn out to be equivalent, and was surprised to find
out that they are not.
}
Ours is more general, as will become apparent in Section \ref{traditional}.

 The more usual notation is to choose a fixed unit vector
$\xi$ in $\cala$, and define 
\beq
\lbl{eq74.15}
U(\phi_i \tnr \xi ) = \phi_i \tnr X_i \q.
\eeq
That may give the impression  that $U$ might later be defined on all of 
$\cals \tnr \cala$. We emphasize that it is only defined on 
the subspace spanned by 
 all $\phi_i \tnr \xi$ , and an extension to all
of $\cals \tnr \cala$ is typically not considered in the literature. 
Araki and Yanase \cite{a/y} then consider
an operator $L : \cals \tnr \cala \rightarrow \cals \tnr \cala $ 
and state as a hypothesis 
that $L$ commutes with $U$:
\beq
\lbl{eq74.20}
UL = LU \q . 
\eeq
But this makes no sense unless $U$ is defined on the range of $L$, 
which, for arbitrary $L$, 
might be expected to contain vectors not of the form $\phi \tnr \xi$.  
There are various ways to circumvent this,
but in general confusion seems inevitable because the reader has no way 
to know which circumvention the authors intended.  
We present one such circumvention below.

Our first task is to fix on an interpretation for \re{eq74.20}.  
The starting point will be to consider $U$ as a map from $\cals$
into $\ts$:  
$$
U : \ \cals \rightarrow \ts \q.
$$
(This accounts for the difference  between the formulation
below and the ``standard'' von Neumann model.) 

If $U: \cals \rightarrow \ts$, 
then the $L$'s on the left and right of \re{eq74.20} must be different,
The domain of the right side is the domain of $U$, namely $\cals$.
Hence the domain of the left side, which is the domain of the left-side $L$, 
must also be $\cals$.  But the domain of the right-side $L$ has to include
the range of $U$, which is a subspace of $\ts$.  The point is that the
domains of the left-side and right-side $L$'s are in different Hilbert
spaces, so the left-side $L$ cannot be the same as the right-side $L$.  
Therefore, we should use different symbols for the two.

Araki and Yanase \cite{a/y} refer to $L$ as a conserved quantity,%
\footnote{
This is a different usage than our usage of ``conserved'' 
to refer exclusively to a quantity which is not changed by a 
{\em measurement}.  Beyond this paragraph, we shall never use
``conserved'' in the Araki-Yanase sense. 
}
which
they define as one which satisfies \re{eq74.20} (their equation (2.6)).
For ease of language, we are going to call this ``conserved'' quantity 
``energy'', with the understanding that it could well be something else
such as spin.  Instead of $L$, we shall use the symbol $H$ for energy,
with an appropriate subscript to indicate to which system the $H$ refers,
$H_\cals$ for the energy operator on $\cals$, 
$H_\ts$ for the energy operator on $\ts$ and $H_\cala$ for the energy 
operator on $H_\cala$.  
Then equation \re{eq74.20} reads:
\beq
\lbl{eq74.25}
UH_\cals = H_\ts U
\q.
\eeq
In the mathematical literature, equation \re{eq74.25}
would be verbalized by saying that $U$ {\em intertwines}  $H_\cals$
and $H_\ts$ instead of saying that {\em $H$ commutes with $U$}. 

This implies that the range of $H_\cals$ is contained in the domain of $U$,
namely $\cals$, so 
$$
H_\cals : \cals \rightarrow \cals  \q. 
$$
Establishing the domain and codomain of $H_\ts$ is a bit trickier.
From the right side of \re{eq74.25}, the domain of $H_\ts$ must contain 
$U\cals$, which is the span of all $\phi_i \tnr X_i$. 
Again from \re{eq74.25}, this span is invariant under $H_\ts$, 
so there seems no harm in taking the codomain of $H_\ts$
as $U\cals $, or as $\ts$ when convenient:  
$$
H_\ts : \ U\cals \rightarrow U\cals \q \mbox{or} \q 
H_\ts : \ U\cals \rightarrow \ts \q.
$$

The question of fixing the domain and codomain of $H_\ts$ 
might seem  nitpicking, but it arises in the folowing way
in the context of the WAY theorem.  Araki and Yanase \cite{a/y} consider
an $H_\ts$ assumed to be of the form
\beq
\lbl{eq74.26}
H_\ts = H_1 \tnr I + I \tnr H_2 \q.
\eeq
Since the right side need not lie in $U\cals$, to even consider such
an $H_\ts$ for arbitrary $H_1$ and  $H_2$, 
one needs to enlarge its codomain beyond $U \cals$.  
However, such an $H_\ts$ can satisfy \re{eq74.25} only if $H_1$ and $H_2$
are such that the range of $H_\ts$ actually does lie in $U\cals$.  
To consider arbitrary $H_\ts$ satisfying \re{eq74.26}, 
it seems reasonable to take the codomain as $\ts$: 
$$
H_\ts : \ U\cals \rightarrow  \ts \q.
$$
But when we want to emphasize that the range of $H_\ts$ must actually
lie in $U\cals$ when \re{eq74.25} holds, we will write 
$$
H_\ts : \ U\cals  \rightarrow U\cals \q.
$$ 

Under the additivity assumption \re{eq74.26},  
the WAY theorem of Araki and Yanase \cite{a/y} 
concludes that the system observable $\cals $ must commute
with the system energy operator $H_1$,  $H_1 S = S H_1$, but that assumes
their ``standard'' von Neumann measurement model.  Within the present  
setup this conclusion would seem something of a red herring for the
following reason.  

It would be natural to imagine that $H_1$ would
be the energy operator on $H_\cals$ and $H_2$ the energy operator
on $H_\cala$.   If that were the case, then commutation of $H_1$
with $S$ would be equivalent to commutation of $H_1$ with 
the projectors of the measurement $\{ Q_i \}$, which by Theorem \ref{thm1} 
is equivalent to conservation of energy in the system space $H_S$.  
This would indeed seem an interesting conclusion.  

But it is rarely the case that $H_1 = H_\cals$. 
At this point, there is no substitute for an explicit calculation
to convince the reader of this.  
Also, this simple calculation contains the essence of our proof
of WAY-type theorems, and illustrates their essential simplicity.

Consider a two-dimensional system $\cals$ with orthonormal bases
$\{ \phi_1 , \phi_2 \}$ for $\cals$, $\{ X_1, X_2 \}$ for $\cala$, 
and $\{ \phi_1 \tnr X_1, \phi_2 \tnr X_2 \}$
for $U\cals \subset \ts$, and a system observable $S = \sum_i \lambda_i P_{\phi_i}$.  
All matrices will be written with respect to whichever of 
these bases is relevant.

Consider arbitrary $H_1$ and $H_2$ with matrices:
$$ 
 H_1 =  
\left[ 
\begin{array}{ll}
h_{11} & h_{12} \\
h_{21} & h_{22}
\end{array}
\right]
 \q \mbox{and} \q 
 H_2 =  
\left[ 
\begin{array}{ll}
k_{11} & k_{12} \\
k_{21} & k_{22}
\end{array}
\right]
\q.
$$
We are going to observe that the necessity for 
$H_\ts = H_1 \tnr I + I \tnr H_2$
to hold invariant the range of $U$, which is the span of $\{ \phi_1 \tnr X_1,
\phi_2 \tnr X_2 \}$, constrains (the matrices of) $H_1$ and $H_2$ to
be diagonal.  
We have
$$
H(\phi_1 \tnr X_1) = h_{11} \phi_1 \tnr X_1 + h_{21}\phi_2 \tnr X_1 
+ k_{11} \phi_1 \tnr X_1 + k_{21} \phi_1 \tnr X_2  
$$
Projecting onto the span of $\{ \phi_i \tnr X_i \}^2_{i=1}$ shows that
$$
h_{21} = 0 = k_{21} \q.
$$
Considering similarly $H(\phi_2 \tnr X_2)$ yields $h_{12} = 0 = k_{12}$,
so $H_1$ and $H_2$ must be diagonal, which implies that 
$H_1$ commutes with $S$.  
Now
$$
H_\ts = H_1 \tnr I + I \tnr H_2 = 
\left[ 
\begin{array}{ll}
h_{11} + k_{11} & 0 \\
0 & h_{22} + k_{22} \\
\end{array} 
\right]
\q,
$$
where the matrix is with respect to the basis $\{ \phi_i \tnr X_i \}$.
The relation $UH_\cals = H_\ts U$ implies that the matrix of $H_\cals$
with respect to $ \{ \phi_1, \phi_2 \}$ is the same as the matrix
of $H_\ts$ with respect to $ \{ \phi_1 \tnr X_1 , \phi_2 \tnr X_2 \}$: 
$$
H_\cals = 
\left[ 
\begin{array}{ll}
h_{11} + k_{11}& 0 \\
0 & h_{22} + k_{22} \\
\end{array} 
\right]
\q \neq H_1 \q.
$$
This shows explicitly that 
$H_\cals$ is unitarily equivalent to $H_\ts | U\cals$, not to $H_1$
as one might imagine.  By the symmetry of the situation, all three of 
$H_\cala, H_\ts |U\cals , H_\cala$ are unitarily equivalent. 

Before continuing, we point out how the essence of the WAY-type theorems,
both ours and that of Araki and Yanase \cite{a/y}, is revealed by 
the above calculation. Our assumed intertwining relation \re{eq74.25}
requires that $H$ hold invariant the range of $U$, which is 
the span of $\{ \phi_1 \tnr X_1,  \phi_2 \tnr X_2 \}.$  
But only very special
$H$ of the form $H = H_1\tnr I  + I \tnr H_2$ can hold invariant this span.

The system energy $H_\cals$ has physical meaning,
but the physical meaning of $H_1$, if any, seems unclear.
And if the physical meaning of $H_1$ is unclear, the import of
the conclusion  that $H_1$ commutes with the system
observable $\cals$, seems even more obscure.  (That conclusion would
be the conclusion of the WAY theorem if the Araki/Yanase setup were 
the same as ours.)  The conclusion that we want is 
that $S$ commutes with $H_\cals$, not $H_1$.  Fortunately, 
we shall see that this 
 desired conclusion does hold, not just for this example but in
general, along with the conclusion that $S$ commutes with $H_1$.   

The reader may wonder if we are merely playing with words in calling 
$H_\cals$ the system energy operator instead of $H_1$, but a little
reflection will dispel this worry.  We started with the system $S$
with energy operator named $H_\cals$.  The system $\ts$ with corresponding
energy operator was just a mathemetical construction.

\subsection{WAY-type theorems in our setup}
\lbl{weakthm}
This subsection will give simple proofs  of variants
of the  WAY theorem in our setup.
The results will be stated in the generality in which  they
have been proved,
%
but to simplify the notation, 
the proofs will assume that the eigenspaces of the system observable
are one-dimensional.  All the important ideas of the proof are 
present in this case.   
The notationally complicated 
proofs for  eigenspaces of possibly greater dimension are relegated
to an appendix. 

For the convenience of skimming readers, we first summarize 
the notation of the preceding sections and introduce two new notations.    
The WAY-type theorems and related results will be mainly consequences of the
general setup developed in Subsection \ref{interpretation}.  To specify
it completely in a theorem's hypotheses would result in an excessively
cumbersome statement. 

Recall that the system observable $S$ has distinct eigenvalues $\lambda_i$
and that the spectral decomposition of $S$ is 
$$
S = \sum_k \lambda_k Q_k
$$
where the $Q_k$ are orthogonal projectors (so that $\{ Q_k \}$ is 
a measurement).    

Let $\{ \phi_i \}_{i\in I}$ be an orthonormal basis for $\cals$, 
where $I$ is some index set, and  
such that for each $k$, the range of $Q_k$ is spanned by some collection
of the $\phi_i$. Thus when $S$ has non-degenerate eigenvalues (i.e., 
all the $Q_k$ are one-dimensional), the notation can be chosen
so that  $Q_k = P_{\phi_k}$, and we assume this choice. 
Let $\{ X_i \}_{i \in I} $ be an orthonormal basis for $\cala$. 

Denote by $U$ the unique isometry $U: \cals \rightarrow \ts$ satisfying 
\beq 
\lbl{eq78}
U \phi_i = \phi_i \tnr X_i \q \mbox{for all $i$.}  
\eeq 
Similarly define an
isometry $V: \cala \rightarrow \ts$ by $VX_i = \phi_i \tnr X_i$.  
Define an ``apparatus observable'' 
$$
A := \sum_k \lambda_k P_k
\q,
$$
where $P_k$ is the projector on $\cala$ with range spanned by the $X_i$ for 
which $\phi_i \in \mbox{Range $Q_k$} $.  Under the simplifying assumption
that $Q_k = P_{\phi_k}$, we have $P_k = P_{X_k}$. 



Let 
$$
H_\cals : \cals \rightarrow \cals\q, \q H_\cala : \cala \rightarrow \cala
\q , \q H_\ts : U\cals \rightarrow \ts \q. 
$$
We call these ``energy operators'' on their respective spaces, 
but they could be arbitrary Hermitian operators. 
We say that the measurement $\{Q_i \}$ {\em conserves energy} on $\cals$
if its operators commute with $H_{\cals}$ :  $Q_i H_\cals = H_\cals Q_i$
for all $i$.  The same language applies to $\ts$ and $\cala$ with
$\{ Q_i \}, H_\cals$ replaced by $\{ UQ_i U^\dag \}, H_\ts$ for $\ts$ 
and 
$\{ P_i \}, H_\cala $ for $\cala$. 

The following proposition is little more than a tautology which systemizes 
the facts which we will need to prove the conclusion of the WAY theorem
for the setup of Subsection \ref{interpretation}  (which is similar but
not identical to
the setup of Araki and Yanase \cite{a/y} ).  

\begin{proposition}
\lbl{weakwaythm}
Assume the general setup just described and that $U H_\cals = H_\ts U$.  Then  
the following are equivalent:
\begin{description}
\item[{\rm (i)}]
$H_\cals$ commutes with $S$:\ \  $S H_\cals = S H_\cals$ ; 
\item[${\rm(i)}^\prime$]
The measurement $\{ Q_i \} $ conserves energy on $\cals$;
\item[{\rm (ii) }]
$H_\ts$ commutes with $USU^\dag$ :\ \  $(USU^\dag) H_\ts = H_\ts (USU^\dag) $; 
\item[${\rm(ii)}^\prime$]
The measurement $\{ UQ_iU^\dag \}$ on the range of $U$ 
conserves energy on $U\cals$;
\item[\rm (iii) ] 
$H_\cala$ commutes with $A$ : \ \ $AH_\cala = H_\cala A$ ; 
\item[${\rm(iii)}^\prime$]
The measurement $\{ P_i \}$ conserves energy on $\cala$. 
\end{description} 
\end{proposition}
{\bf Proof:}
This proof does not require the simplifying assumption that $Q_i = P_{\phi_i}$. 
The equivalence of the various items and their primed versions 
(e.g., (i) and $\rm (i)^\prime$) is immediate
from the spectral theorem, part of which states that an operator 
commutes with $S$ if and only if it commutes with all of the spectral
projectors $Q_i$ for $S$.

Since the domain of $H_\ts$ is $ U \cals$, 
to show that $ UQ_iU^\dag $ commutes with $H_\ts$
when $H_\cals$ commutes with $Q_i$, 
it is sufficient to show that $(H_\ts UQ_i U^\dag)U =  ( UQ_iU^\dag H_\ts)U $. 
We have, using $U^\dag U = I$ and $H_\cals Q_i = Q_i H_\cals$,  
\begin{eqnarray*} 
[H_\ts (U Q_i U^\dag)] U 
 &=& U H_\cals Q_i \\
&=& U Q_i H_\cals \\ 
&=& (U Q_i U^\dag)(UH_\cals) \\
&=& [(U Q_i U^\dag)H_\ts] U 
\q.
\end{eqnarray*}
For the converse, that $Q_i H_\cals = H_\cals Q_i$ when 
$ (UQ_iU^\dag ) H_\ts = H_\ts (UQ_i U^\dag)$, first note that using 
$U^\dag U = I$, 
$UH_\cals = H_\ts U$ implies that $H_\cals = U^\dag H_\ts U$.  Hence
\begin{eqnarray*}
 H_\cals Q_i &=& U^\dag H_\ts U Q_i  \\
&=& U^\dag H_\ts (UQ_i U^\dag) U \\
&=& U^\dag (UQ_i U^\dag) H_\ts U \\
&=& Q_i U^\dag H_\ts U \\
&=& Q_i H_\cals \q.
\end{eqnarray*} 

We have shown that  (i), $\rm (i)^\prime$, (ii), and $\rm (ii)^\prime$ are
equivalent.  The equivalence of (iii), and $ \rm (iii)^\prime$ with the rest
follows similarly from  the symmetry of the setup.
\boxx

Next we obtain WAY-type theorems from the Proposition. 
Various hypotheses on the form of $H_\ts$, such as the Araki/Yanase 
assumption that it is of the form $H_\ts = H_1 \tnr I + I \tnr H_2 $,
are easily seen to imply that $H_\ts$ commutes with the $U Q_i U^\dag$,
which from the proposition implies that energy is conserved not only 
in $\ts$, but also in 
$\cals$ and $\cala.$.

In my view, this is the physically relevant conclusion that one wants,
as discussed in Subsection \ref{interpretation}. However, 
it is not the form of the  conclusion of Araki/Yanase's WAY theorem
\cite{a/y}, that $S$  commutes with $H_1$.  For comparison and 
completeness, we obtain the latter
also.


Say that $H_\ts : \ts \rightarrow \ts$ is {\em diagonal} with respect 
to the projectors 
$\{ U Q_i U^\dag \}$ if $(U Q_j U^\dag ) H_\ts (U Q_i U^\dag) = 0 $  
for all $i$  and all $j \neq i$, with a similar meaning for  ``$H_\cals$
is diagonal with respect to $\{ Q_i \}$'', etc.   
When $H_\ts$ holds invariant the range of $U$, (as it does when
$H_\ts U = U  H_\cals$) 
 this implies that 
$H_\ts$ commutes with all $U Q_i U^\dag$ and conversely.

The next result looks like the WAY theorem expressed in our
setup, though (in the context of non-degenerate eigenvalues for the 
system observable) 
it is actually more general, as shown in Section 
\ref{traditional}. 
Its statement assumes the notation of the preceding discussion.  
However, we summarize it first for the benefit of skimming readers 
who may want to get a feel for the result in order to decide whether 
to read further.  

The object of interest is a quantum (``system'')
observable $S$ on a Hilbert space $\cals$ which is to be measured.  
The measurement apparatus is an observable $A$ on a Hilbert space $A$,
and the total system-apparatus Hilbert space is $\ts$.  On each 
of these three Hilbert spaces is defined an energy
operator denoted $H_\cals$ for $\cals$, $H_\cala$ for $\cala$, 
and $H_\ts$ for $\ts$.  These are related by an isometry $U$
which intertwines $H_\cals$ and $H_\ts: \ UH_\cals = H_\ts U$ 
and embeds $\cals$ into $\ts$.  The measurement is carried out in $\ts$,
but the result is translated back into $\cals$ via the identification
furnished by $U$.   
\begin{theorem}[WAY-type theorem with Yanase-type condition proved.]
\lbl{cor3} 
Suppose that $H_\ts$ is of the form 
$$
H_\ts = H_1 \tnr I + I \tnr H_2 \q
$$
and that $UH_\cals = H_\ts U$. Then the system observable $S$
commutes with both the system energy operator $H_S$ and $H_1$,  
and the apparatus observable $A$ commutes with both the apparatus
energy operator $H_A$ and $H_2$. 
\end{theorem}
{\bf Proof} (for $S$ with one-dimensional eigenspaces): 
Though stated in general, for simplicity and clarity this proof will
be given under the simplifying assumption that $Q_k = P_{\phi_k}$
and $P_k = P_{X_k}$.  The general proof is relegated to the Appendix.

To show that $H_1$ commutes with $S$, we exploit the fact that because of 
$H_\ts U = U H_S$, $H_\ts$ holds invariant the range of $U$, which is spanned
by the $\phi_i \tnr X_i$. 
Suppose for some $k \neq j$, $ \lb \phi_k , H_1 \phi_j \rb \neq 0$.
Then 
$$
\lb \phi_k \tnr X_j , H_\ts(\phi_j \tnr X_j) \rb 
=
\lb \phi_k, H_1 \phi_j \rb \cdot 1 + 0 \cdot \lb X_j , H_2 X_j \rb \neq 0 
$$
shows that $H_\ts$ does not hold invariant the span of all $\phi_i \tnr X_i$. 
If it did, we would have $H_\ts ( \phi_j \tnr X_j) = 
\sum_i c_i \phi_i \tnr X_i$ for some scalars $c_i$, 
and consequently, $\lb \phi_k \tnr X_j , H_\ts (\phi_j \tnr X_j) \rb = 0$.  
The proof that $H_2$ commutes with the apparatus observable $A$ 
is the same, using the system-apparatus symmetry of the setup.

Thus 
$$ H_1 \phi_i  = d_i \phi_i \q \mbox{and} \q H_2 X_i = b_i X_i \q 
\mbox{for some scalars $d_i, b_i$ and all $i$,}  
$$
and consequently
$$
H(\phi_i \tnr X_i) = (d_i + b_i) \phi_i \tnr X_i \q.
$$
The relation $UH_\cals = H_\ts U $ shows that  
 $U$ implements a unitary equivalence between $H_\cals$ and 
$H_\ts |\, \mbox{Range $U$}.$  Under this equivalence, the basis 
$\{ \phi_i \}$ for $\cals$ 
goes over into the basis $\{ \phi_i \tnr X_i \}$ for $\ts$.  Hence the matrix
of $H_\cals$ with respect to $\{ \phi_i \}$ is the same as the matrix
of $H_\ts$ with respect to $\{ \phi_i \tnr X_i \}$, namely  the diagonal
matrix diag $ ( d_i + b_i)$.  The matrix of $S$ with respect to 
$\{ \phi_i \}$ is also diagonal, namely diag $(\lambda_i ).$
Hence $H_\cals$ commutes with $S$.%
\footnote{
A slightly simpler proof is given in the Appendix for the general
case of possibly degenerate eigenvalues, written after and independently
of the above.   
I decided not to change the above proof because 
though slightly longer, it demands less of the reader, and
its observations could be helpful for the next Section \ref{traditional}.
} 
Similarly, $A$ commutes with
$H_2$, which is called the ``Yanase condition'' (see below),
and also with $H_\cala$. \ \  \boxx

The so-called ``Yanase condition'' requires some explanation.
A year after the Araki/Yanase WAY theorem \cite{a/y} was published,
Yanase published \cite{yanase} which seems to adjoin in some way
to the WAY theorem
the  condition that $H_2$ commute with the apparatus observable $A$.
This has become known as the ``Yanase condition''.
Yanase's language is obscure to me, and seemingly to other authors.
There may be various interpretations, but there seems substantial
agreement that the Yanase condition is physically desirable.

The above proofs are valid under hypotheses on the form of $H_\ts$
considerably more general than the Araki/Yanase hypothesis 
$H_\ts = H_1 \tnr I + I \tnr H_2$
stated, but more general formulations make the hypotheses too cumbersome
and obscure the simplicity of the proofs.  We indicate here some 
typical generalizations.

	If $D_1$ and $D_2$ are diagonal  Hermitian operators 
(with respect to $\{ Q_i \}$ and $\{P_i\}$ respectively,
then the proof of Theorem \ref{cor3} goes through 
with hypothesis
\beq
\lbl{eq80}
H_\ts = H_1 \tnr D_2 + D_1 \tnr H_2 \q.
\eeq
 
If $D_1$ is a Hermitian operator which is diagonal with respect to $\{Q_i \}$
and
\beq
\lbl{eq83}
H_\ts = D_1 \tnr (V^\dag U)D_1 (V^\dag U)^\dag \q,
\eeq 
the easy part of the proof
 of Theorem \ref{cor3} establishes its conclusion. 
(Again, $U$ implements a unitary equivalence between $H_\cals$ and
the operator $H_\ts$, which is obviously diagonal with respect 
to $\{\phi_i \tnr X_i \}$.)
The above statements involving forms \re{eq80} and \re{eq83}
assume non-degenerate eigenvalues (i.e., the $\{Q_i \}$ are one-dimensional). 
I have not examined more general cases. 
\section{Comparison with the traditional approach}
\lbl{traditional} 
Again, we denote the system Hilbert space by $\cals$ and the apparatus
space by $\cala$, with $\ts$ the system-apparatus space.  
For simplicity, the discussion will assume that the spectral projections
of the system observable have one-dimensional range.  
Let $\{ \phi_i \}$ be an orthonormal basis for $\cals$ and $\{ X_i \}$
an orthonormal basis for $\cala$.  Assume a system observable $S$
of the form
$$
S = \sum_i \lambda_i P_{\phi_i} \q  \mbox{with distinct $\lambda_i$.}
$$ 
Let $\xi$ denote some distinguished unit vector in $\cala$ and
$[\xi]$ the one-dimensional subspace that it spans. 
Define 
\beq
\lbl{eq700}
U :  \cals \tnr [\xi] \rightarrow \ts  
\eeq
to be the unique isometry which satisfies
\beq
\lbl{eq710}
U(\phi_i \tnr \xi) = \phi_i \tnr X_i \q \mbox{for all $i$}.
\eeq 

Since $\cals \tnr [\xi]$ is naturally identified 
with $\cals$ via the map $\phi \tnr \xi  \mapsto \phi$, I formulated
the approach of Section \ref{waytype} in the expectation that
the results would be equivalent to those of the traditional approach.
I was surprised and initially puzzled 
when they turned out to differ significantly.
This will be discussed in more detail below.
We shall also derive the original WAY theorem (for the special case
of non-degenerate system observable eigenvalues) as formulated
by  Araki and Yanase from the Section \ref{waytype} results.

Let $H$ denote the energy operator on $\ts$, previously called $H_\ts$.
We change the name to avoid confusion with the previous approach and
for easier comparison with the approach and notation
of Araki/Yanase.%
\footnote{However, Araki and Yanase use $L$ for what they call the
``conserved'' quantity instead of our $H$.
}
The traditional approach allows us to write the equation
\beq
\lbl{eq720}
HU = UH
\eeq
with some hope of giving it meaning, since all operators in it are 
defined on some subspace of $\ts$.  However, since 
$\ U : \, \cals \tnr [\xi] \rightarrow \ts$  is never extended to $\ts$,
the traditional approach  seems to me algebraically unnatural.
The approach of Subsection \ref{interpretation} turns out to be 
actually, not just cosmetically, more general.

For \re{eq720} to be meaningful,
$U$, which so far is only defined on the subpace $\cals \tnr [\xi]$,
must hold invariant the range of $H$.  In particular, the range
of $H$ must be in the domain of $U$, which is $\cals \tnr [\xi ]$.  

Araki and Yanase assume that $H$ is of the form
$$
H = H_1 \tnr I + I \tnr H_2 
\q,
$$  
with $H_1$ the energy operator on $\cals$ and $H_2$ the energy operator
on $\cala$. 
We shall see that this that this forces $H_2$ to be a multiple 
of the identity operator.  We leave it to the reader to decide whether
this is physically reasonable.

We have 
$$
H (\phi \tnr \xi) = (H_1 \phi ) \tnr \xi + \phi \tnr (H_2 \xi)
\q.
$$
For this to be in the domain of $U$ (so that we can write $UH = HU$), 
we must have 
\beq
\lbl{eq730}
H_2 \xi = a \xi \q \mbox{for some scalar $a$.}
\eeq
We do not exclude the case $a = 0$.  Finally, we have
\beq
\lbl{eq800}
H(\phi \tnr \xi ) = (H_1 \phi) \tnr  \xi + a\phi \tnr \xi  
\eeq 

Before continuing, we must deal with a notational problem.  
In Section \ref{waytype},  we started with an observable $S$ on an 
abstract Hilbert space $\cals$ and 
an energy operator $H_\cals \ : \cals \rightarrow \cals$ on the same space.

In the traditional model which we are analyzing, $\cals$ is effectively
replaced by $\calsprime := \cals \tnr [\xi] \subset \ts $ by identifying  
 $\cals$  with  
$\calsprime$ via the unitary map 
$$ 
W \ : \cals \rightarrow \calsprime,  \q\q \phi \mapsto \phi \tnr \xi \q. 
$$
In our discussion of the traditional model, $\calsprime$ will take the
place of the $\cals$ in the Section \ref{waytype} discussions. 

Under the unitary identification $W$, an energy operator $H_\cals$
on $\cals$ corresponds to 
\beq
\lbl{eq810}
W H_\cals W^\dag : \  \calsprime \rightarrow \calsprime 
\eeq    
on $\calsprime$.
So, $ W H_\cals W^\dag ( \phi  \tnr  \xi  ) = ( H_\cals \phi)  \tnr  \xi .$ 
However, the traditional approach does not explicitly specify 
an energy operator on $\cals$ or $\calsprime$.  To continue to compare
the approaches, we must decide which operator on $\calsprime$ to 
take as the analog of the Section \ref{waytype} energy operator 
$H_\cals$.  A natural
assumption is that $H_1$ is to be taken as $H_\cals$, the energy operator
on the present $\cals$, and we shall make that assumption:  
$$
H_\cals := H_1 \q.
$$
           
But now  we have {\em two} ``natural''  candidates for 
an energy operator on $\calsprime$,
$ WH_\cals W^\dag = W H_1 W^\dag,$   and the restriction
of $H$ to the invariant subspase $\calsprime$, denoted $H | \calsprime$.    
Equation \re{eq800} shows that these are not necessarily the same: 
$$
H|\calsprime =  WH_1 W^\dag + aI \q . 
$$
However, the difference is physically insignificant because an energy operator 
is defined only up to an arbitrary additive constant multiple of the identity.
It also turns out to be mathematically insignificant:  both ``natural''
definitions lead to the  conclusion of the WAY theorem. 


We can make any definition that we 
want for the energy operator $H_\calsprime$ on $\calsprime$, but 
it is comforting that for our purposes, it will be irrelevant 
which of the two
``natural'' definitions we use.   
We shall define
\beq 
\lbl{eq815}
H_\calsprime := H|\calsprime \q.
\eeq
We also define a new system operator $\Sprime$ on $\calsprime$ to 
be the old $S$ transferred to $\calsprime$ via the identification $W$:
$$
\Sprime := WSW^\dag \q.
$$

Now we are in precisely 
the situation considered in Section 
\ref{waytype} with its system space $\cals$ replaced by $\calsprime$,  
its system observable $S$ by 
$\Sprime$, its system energy operator $H_\cals$
 by $H_\calsprime$, and its $U$ by the $U$ defined  above
by \re{eq710}, so its analysis and the theorems proved there
apply directly. We do not rename $U$  because whether the 
$U$ of Section \ref{waytype} or the $U$ of \re{eq710} is meant will always
be clear from the context.  (It will always be the $U$ defined by 
\re{eq710}.)

The hypothesis $UH_S = H_\ts U$ of Theorem \ref{cor3}
here reads 
$$
U H_\calsprime = H U \q. 
$$
Since the domain of both sides is $\calsprime$, that is tautologically
euivalent to 
$$
UH_\calsprime | \calsprime = HU | \calsprime \q, 
$$
which does hold under the hypothesis $UH = HU$ of the Araki/Yanase WAY
theorem \cite{a/y} because 
$$
UH_\calsprime | \calsprime  = UH|\calsprime = HU | \calsprime  \q. 
$$ 
Now Theorem \ref{cor3} implies that 
the new system observable
$S^\prime := W S W^\dag$ 
commutes with $WH_1 W^\dag $ ($H_1$ transferred to $\calsprime$),
and hence $S$ commutes with $H_1$. 
This proves the WAY theorem of Araki and Yanase \cite{a/y} for the 
special case of nondegenerate eigenvalues, but we can say more. 

We shall next observe that necessarily, $H_2 = aI$.
We proved above that $H_1$ commutes with the system observable $S$,
for which  $\phi_i$ are eigenvectors corresponding to distinct eigenvalues.
That implies that 
\beq
\lbl{eq835}
H_1 \phi_i = d_i \phi_i \q \mbox{for sone scalars $d_i$ and all $i$.} 
\eeq
From $UH = HU$, it follows that $H = H_1 \tnr I + I \tnr H_2$ holds 
invariant the range of $U$, which is the span of $\{ \phi_i \tnr X_i \}$.
From $H_i \phi_i  = d_i \phi_i$, it follows that $H_1 \tnr I $ holds
invariant this range.  Hence $I \tnr H_2$ must also hold invariant 
the span of $\{ \phi_i \tnr X_i \}$.  But from the Parseval equality,
$$
(I \tnr H_2 )(\phi_i \tnr X_i ) = \phi_i \tnr H_2 X_i 
= \sum_j \lb X_j , H_2 X_i \rb \phi_i \tnr X_j.
$$
For $j \neq i$, $\phi_i \tnr X_j$ is orthogonal to all $\phi_k \tnr X_k$,
as well as all $\phi_i \tnr X_m$ for $m \neq j$,
so $\lb X_j , H_2 X_i \rb = 0 $ for $ j \neq i $, which shows (again by 
Parseval) that 
$$
H_2 X_i  = b_i X_i \q \mbox{with $b_i = \lb X_i , H_2 X_i \rb$.}
$$

Finally, $U$ implements a unitary equivalence between 
$H|\calsprime = H_1 \tnr I + I \tnr aI$ and $H| \, \mbox{Range $U$},$
so the eigenvalues of the former, namely $d_i + a$ must equal the eigenvalues
of the latter, namely $d_i + b_i$.  Hence $b_i = a$ for all $i$.
In more detail,  
$$
H (\phi_i \tnr \xi) = H_1 \phi_i \tnr \xi + \phi \tnr H_2 \xi = 
(d_i + a) \phi_i \tnr \xi , \q \mbox{and}
$$ 
$$
H (\phi_i \tnr X_i ) = H_1 \phi_i \tnr X_i + \phi_i \tnr H_2 X_i 
= (d_i + b_i ) \phi_i \tnr X_i .
$$
Since $UH = HU$,
\begin{eqnarray*}
\lefteqn{ (d_i + a ) \phi_i \tnr X_i =  (d_i + a )  U(\phi_i \tnr \xi ) 
= U((d_i+a_i)(\phi_i \tnr \xi))} \\ 
&=& UH(\phi_i \tnr \xi ) 
= HU (\phi_i \tnr \xi) = H(\phi_i \tnr X_i ) = (d_i + b_i) \phi_i \tnr X_i,
\end{eqnarray*}
whence $d_i + b_i = d_i + a $ for all $k$, so $b_i = a$ and 
\beq 
\lbl{eq845}
H_2 = aI \q.
\eeq
Of course, this implies the Yanase condition (that the apparatus
observable with eigenvectors $\{X_i\}$ commutes with $H_2$).

I was surprised by the conclusion $H_2 = aI$ because it seems 
so unphysical.
How are we to understand this?  Why did $H_2 = aI$ not already appear in the
Section \ref{waytype} analysis?  This bothered me because I was worried 
that there might be an error, and I wanted to understand $H_2 = aI$ 
independently of the detailed analysis.  
The way of looking at the situation that
convinced me that there probably is no error appears just below
for the benefit of the reader who may be similarly uneasy.

Once perceived, the reason for the difference is easy to understand.
There was no $a$ in the Section \ref{waytype} analysis.  It entered 
the present analysis at
$$
H_\calsprime (\phi \tnr \xi)  :=  H(\phi \tnr \xi) 
= (H_1 \phi) \tnr \xi + \phi \tnr H_2 \xi 
= (H_1 \phi) \tnr \xi + \phi \tnr a \xi,
$$
which arose because $H_2 \xi = a\xi$ was forced 
in order to make $UH = HU$ well defined.
This makes the restriction of $I \tnr H_2$ to $\calsprime$ equal to $aI$:
$(I \tnr H_2)|\calsprime = aI$.  

Independently of our analysis,
the Araki/Yanase WAY theorem implies that $H_1 \phi_i = d_i \phi_i$ for
some scalars $d_i$.  It follows that 
$U$ intertwines $(H_1 \tnr I)| \calsprime$ and 
$(H_1 \tnr I)| \mbox {Range $U$}:$  
$$
U(H_1 \tnr I) (\phi_i \tnr \xi ) = (d_i \phi_i) \tnr X_i = 
(H_1 \tnr I ) U ( \phi_i \tnr \xi ) .
$$ 

Also, from $UH = HU$, $U$ intertwines
$H|\calsprime$ and $H| \mbox{Range $U$.}$  Hence, from
$H = H_1 \tnr I + I \tnr H_2$,  $U$ also intertwines
$(I\tnr H_2)|\calsprime$ and $(I \tnr H_2) | \mbox{ Range $U$.} $    
Since $(I \tnr H_2)$ is $aI$ on $\calsprime$, it must also be $aI$ 
on Range $U$, i.e., $H_2 = aI$.

The following theorem summarizes. The first paragraph of hypotheses 
merely summarizes the setup just described. 
\begin{theorem}[Extension of WAY theorem for non-degenerate eigenvalues] 
\lbl{thm3}  
Let $S = \sum_i \lambda_i P_{\phi_i} $ be 
 an observable on a Hilbert space $\cals$, where $\{ \phi_i \}$ is 
an orthonormal basis for $\cals$, and the $\lambda_i$ are 
real scalars.  Let $X_i$ be an orthonormal basis for
a Hilbert space $\cala$, 
where $i$ runs over the same index set as for 
$\phi_i$.  Let $\xi$ be a unit vector in $\cala$. 
Define an isometry $U : \cals \tnr [\xi ] \rightarrow \ts$ as the 
unique isometry satisfying 
$ U(\phi_i \tnr \xi ) := \phi_i \tnr X_i $.  

Let $H$ be an operator on $\ts$ of the form  $H = H_1 \tnr I + I \tnr H_2$,
where $H_1 : \ \cals \rightarrow \cals $ and $ H_2: \ \cala \rightarrow 
\cala $.
Assume that $HU$ commutes with $UH$ in the sense that for any  
$\phi_i \tnr \xi$, 
$HU ( \phi_i \tnr \xi )= UH ( \phi_i \tnr \xi) .$  

Then 
\begin{description} 
\item[\rm (i)]  $H_1$ commutes with $S$; equivalently, 
$H_1 \phi_i = d_i \phi_i$ for some constants $d_i$;
\item[\rm (ii)]  The measurement $\{ P_{\phi_i} \}$ defined by the
spectral projectors $P_{\phi_i}$ for $\cals$ conserves energy, as defined by 
the energy operator $H_1$ on \cals; 
\item[\rm (iii)] Necessarily, $H_2 \xi = a \xi$ for some constant $a$,
and $H_2 = aI$. 
\end{description}
\end{theorem}
This was proved above.  Condition (iii) implies the Yanase condition.  
The proof of Theorem \ref{thm3} does not require that the  $\lambda_i$ be 
distinct. 
I believe that the generalization to
degenerate eigenvalues should be routine, 
but I have not worked out the details.  
\section{Return to the original conundrum}
\lbl{return}
\subsection{The conundrum}
\lbl{conundrum}
The original motivation for studying the WAY theorem was to figure out
which hyothesis might be responsible for its hard-to-believe conclusion
that all (discrete) observables (which can be exactly measured) commute
with the energy operator.  The condition that the observable be discrete
is probably not the origin of the problem; quite likely a similar theorem
could be proved for observables like position and momentum with a
``continuous'' spectrum.  Or, since real measurements cannot be made with
arbitrary precision, one could probably construct a quantum mechanics
for which all observables are discrete.  

The condition that the discrete observables be ``exactly'' measurable
(in a sense defined in \cite{wigner, a/y} but not discussed in the 
present paper) also seems relatively harmless.  Doesn't a Stern-Gerlach
apparatus ``exactly'' measure whether a particle's spin in a given
direction is ``up'' or ``down''?  The particle leaves the apparatus
going in just one of two possible directions which are easily distinguishable.

The other two hypotheses of the  WAY theorem are the assumption of a
von Neumann type measurement model throughout, and the assumption that
the energy observable $H$ is ``additive'': $H = H_1 \tnr I + I \tnr H_2$.  
The assumption that $H$ is additive can certainly can be questioned.   
For example, suppose the quantum system is an electron, and the 
apparatus a proton.  We would certainly not expect the energy 
operator for a hydrogen atom to be the sum of the energy operators
for a free electron and for a free proton.  (Of course, this is intended as 
a metaphor, not as a mathematically meaningful objection.) 

However, even if additivity seems unlikely in {\em all} situations,
it seems that it could be possible in {\em some} situations. 
And in those situations,  it would seem strange if {\em no} observable which
did not commute with the energy operator could be (exactly) measured, as the 
WAY theorem implies.   That focuses attention to the von Neumann model
as possibly the physically  unrealistic assumption. 
%
%
\subsection{Can the von Neumann model be generalized to imply
conservation of energy in measurements?}

Let us return to the observation of the Introduction that quantum
projective measurements, need not conserve energy.  Since there is 
arguably no principle more pervasive in physics than conservation of
energy, this is certainly unsettling.  A natural way to save the principle 
 is to 
imagine that every measurement involves an interaction of the measured
system with a measuring apparatus, and that energy gained or lost by
the system would be lost or gained by the apparatus.  

The WAY-type theorems show that under the hypotheses of  
Araki and Yanase \cite{a/y}, 
or those of Section \ref{waytype}, such
a resolution is impossible (or unnecessary) because under those hypotheses, 
the system energy is {\em always
conserved} by a measurement.
Measurements which should violate conservation of energy according
to textbook quantum mechanics are simply impossible under the WAY
hypotheses.   If a von Neumann-type measurement model is to be 
retained, then it seems that the hypothesis of additivity of the
energy observable, $H = H_1 + H_2$, should go.

\section{Appendix}
\lbl{appendix}
This appendix indicates how the simplifying assumption of the proofs
of the WAY-type theorems in Subsection \ref{weakthm}, that the system
observable has non-degenerate eigenvalues, can be removed. 
This is just for the reader's convenience because the extensions
will be more or less routine, though annoyingly complicated.  
\\[2ex] 
{\bf Proof of Theorem \ref{cor3}:}
We use the notation of Subsectiion \ref{weakthm}. Let 
$$
\phi_{k,1} ,\  \phi_{k,2} ,\  \ldots ,\  \phi_{k,n_k}
$$
be an orthonormal basis for the spectral subspace of the system 
observable $S$ which is the range of its spectral projection $Q_k$
for eigenvalue $\lambda_k$.  
In other words, $\{ \phi_{k,j} \}^{n_k}_{j=1}$ spans Range $Q_k$.
For convenience, the notation assumes that this range
is finite dimensional, but that is unnecessary. 
Let $ \{ X_{k, j} \} $ be a similarly indexed orthonormal basis for $\cala$,
so that $\{ X_{k,j}\}^{n_k}_{j=1} $ spans the range of the spectral projection
$P_k$ for eigenvalue $\lambda_k$ of the apparatus observable $A$. 

Define $U: \cals \rightarrow \ts$ as the unique isometry satisfying  
$$
U\phi_{k,j} := \phi_{k,j} \tnr X_{k,j} \q \mbox{for all $k$ and 
$1 \leq j \leq n_k$.}
$$
Similarly define $V: \cala \rightarrow \ts$ by
$$ 
VX_{k,j} := \phi_{k,j} \tnr X_{k,j} \q 
\mbox{for all $k$ and $1 \leq j \leq n_k$,}
$$ 

To show that $H_1$ commutes with $S$, suppose for some $\phi_{k,j}$ 
and some $\phi_{m, p} $ with $m \neq k$, 
$\lb \phi_{m,p} , H_1 \phi_{k,j} \rb \neq 0.$  Then 
$$ 
\lb \phi_{m,p} \tnr X_{k,j} \,   , \,  H_\ts (\phi_{k,j} \tnr X_{k,j}) =
 \lb \phi_{m,p}\, , \,  H_1 \phi_{k,j} \rb \cdot 1 + 
0 \cdot \lb X_{k,j}\,  ,\, H_2 X_{k,j} \rb \neq 0
$$
shows that $H_\ts (\phi_{k,j} \tnr X_{k,j} ) $  is not contained in the
range of $U$, which is Span $\{ \phi_{s,t} \tnr X_{s,t} \}$, contrary
to the assumed $H_\ts U = U H_\cals$ .  Hence for all $k$, 
$H_1$ holds invariant  $ \mbox{Range $Q_k$} = $ Span 
$ \{ Q_{k,j} \}^{n_k}_{j=1} $, which for Hermitian $H_1$ 
is equivalent to $H_1 Q_k = Q_k H_1$, and to $H_1 S = S H_1$.  
By the system-apparatus symmetry, the same argument shows that $H_2$ 
commutes with the apparatus observable $A$ and all its spectral projections
$P_k$.



Next we show that $S$ commutes with  $H_\cals$.
Let $\calr_k$ denote the span of all $\phi_{k,j} \tnr X_{k,j}$ for 
$1 \leq j \leq n_j$.
Now  $U H_\cals = H_\ts U$ together with th form of $U$ shows that  
$U$ implements a  unitary equivalence of $H_\cals$ with 
 $H_\ts | \mbox{Range $U$}$, 
which sends $\mbox{Range $Q_k$}$ onto $\calr_k$.  Let 
$\phi_k \in \mbox{Range $Q_k$} $, say $\phi_k = \sum_j c_{k,j} \phi_{k,j}$. 
Let $X_k :=  
\sum_j c_{k,j} X_{k,j}.$, where the $c_{k,j}$ are the same as in the
expansion for $\phi_k$.
Then $U \phi_k = \phi_k \tnr X_k$.

We have
\beq
\lbl{eq945} 
UH_\cals \phi_k =  H_\ts U \phi_k = H_\ts (\phi_k \tnr X_k) = 
 (H_{1}\phi_{k}) \tnr X_{k} 
+  \phi_{k} \tnr H_{2} X_{k} . 
\eeq
We want to show that $H_\cals$ holds Range $Q_k$ invariant.  Since 
$U$ is an isometry, this is the same as showing that 
for $s \neq k$, \re{eq945}
is orthogonal to $U ( \mbox{Range $Q_s$} )  .$ 
That this is so is seen by
taking the inner product of the right side of \re{eq945} with
a typical vector spanning $U( \mbox{Range $Q_s$})$ for $s \neq k$, say
$\phi_{s,t} \tnr X_{s,t}$.
Each of the terms on the rignt side will have a factor 
$\lb X_{s,t} \, , \, X_k \rb = 0 $ or $ \lb \phi_{s,t} \, , \phi_ k \rb = 0$,
so the total result is zero.


Finally, the invariance of Range $Q_k$ under $H_\cals$ shows that
$H_\cals$ commutes with all $Q_k$ and with $S$.  Similarly,
$H_\cala$ commutes with all $P_k$ and $A$.  
This completes the proof of the
stated version of Theorem \ref{cor3} without the simplifying
assumption that the $Q_i$ have one-dimensional ranges.
\  \    \boxx

It should be noted that the statement of the WAY theorem by Araki and 
Yanase in \cite{a/y} is even more general than what we have just proved.  
For that reason,
we continue to call Theorem \ref{cor3} a ``WAY-type'' theorem.  
I know of no obstacle to the proof of the full Araki/Yanase statement
\cite{a/y}, but also I have not thought through what might be involved. 


\end{document}